# ParisTech Review

# Onomastics for Business: can discrimination help development?

*Elian Carsenat* / Founder of NamSor Applied Onomastics / September 17th, 2013

progress and growth    technology and business

*As of today, the main business application of onomastics is naming, or branding: finding the proper name for your company or your product to stand out in the world. Meaningfully, Onoma – the Greek root for name – is also a registered trademark of Nomen, the naming agency founded by Marcel Botton in 1981. Nomen initially licensed one of Roland Moreno's inventions, the Radoteur name generator, and created many distinctive and global brand names such as: Vinci, Clio or Amundi. But once your business has a name, should you forget about onomastics? Not anymore. Globalization, digitalization and the Big Data open new fields to experiment disruptive applications in Sales & Marketing, Communication, HR and Risk Management. Though discriminating names carries a high risk of abuse, it can also drive new, unexpected ways for developing poor areas.*

Our human brain interprets names every day, as we understand a language, as we know a particular culture or region of the world: the likely menu of a restaurant, the industrial sector of a company… even a dog's name might tell you something about its owner. Personal names (first name, last name, a Twitter handle) carry meanings which vary according to one's language and culture, but often form an essential part of one's identity.

**Extracting semantics from names**
How exactly my brain works is not clear even to myself, but what if I could program a computer to extract semantics from names: would it provide valuable business intelligence? Some people in the US think so. The Central Intelligence Agency (CIA) has a long standing experience in extracting intelligence from personal names: back in the 80s they used LAS name recognition software to help identify Russian spies, recognize false identities, track soviet influence. LAS could rely on the CIA to help collect a database with one billion names to calibrate the software. That's about the total world developed population at the time.

After thriving on the surge in US security and foreign intelligence budgets post-9/11, LAS considered diversification and started to address other markets: Marketing, Financial Services Compliance (notably KYC, ie. Know Your Customer). LAS was acquired by IBM in 2006. But to further increase their leadership, in 2011 the US security agencies used the MITRE Corporation to help foster further "innovation in technologies of interest to the federal government. Challenge #1 entailed multicultural name matching—a technology that is a key component of identity matching, which involves measuring the similarity of database records referring to people. Uses include verifying eligibility for Social Security or medical benefits, identifying and reunifying families in disaster relief operations, vetting persons against a travel watch list, and merging or eliminating duplicate records in databases. Person name matching can also be used to improve the accuracy and speed of document searches, social network analysis, and other tasks in which the same person might be referred to by multiple versions or spellings of a name". A name tells more – or something different – than just a nationality of origin. For example, Boston terrorists Tamerlan and Dzhokhar Tsarnaev have names with a -v termination typical of Slavic names (as found in Russia or in Bulgaria) but can be recognized as originally from Caucasus. There was some media report in the aftermath of the bombing that the FBI didn't know Boston bomber travelled to volatile Dagestan region in Russia in 2012 because "his name was misspelled on travel documents". However this information remained unconfirmed and is probably not accurate given the massive US investment in name-matching technology.

In Europe, the legal framework to leverage such tools varies from country to country, but is generally very strict. The directive 95/46/EC of the European Parliament and of the Council of 24 October 1995 on the protection of individuals with regard to the processing of personal data and on the free movement of such data, article 8, states that "Member States shall prohibit the processing of personal data revealing racial or ethnic origin, political opinions, religious or philosophical beliefs, […]" . In principle, this directive applies to Security Agencies as well, however there are exemptions which member states can interpret differently.

By making the distinction between the language 'discriminatory ethnic profiling' rather than the more common 'ethnic profiling' to describe the practice of basing law enforcement decisions solely or mainly on an individual's race, ethnicity or religion the European Union recognizes the need of security forces to understand the complex relationships that exist between nationality, geography, and more subjective concepts such as: ethnic origins, cultural backgrounds, civilisations, religions. How the knowledge might be applied, how the data might be collected remains a matter of national security. The UK and France, for example, are known to have different views on this topic. In any case, what is done in practice by anti-terrorism agencies is not public information.

Security, border control, etc. is a business in its own right. What about other sectors?

**Customer intelligence: business potential and ethical issues**
In Sales & Marketing, onomastics can be used to enrich a customer database with information extracted from names that would not be practically or economically available otherwise. So retailers and luxury brands – especially in food, clothing and cosmetics where ethnicity plays a significant role – can improve customer intelligence and use those insights to better interact through online channels. Echoing concerns expressed by early 20th century John Wanamaker "Half the money I spend on advertising is wasted; the trouble is I don't know which half", companies like L'Oreal that spend several billion dollars a year on communication and advertising continuously try to improve the efficiency of their targeting.

Let us look at a more sophisticated example, for example public–private partnership (PPP) projects in mining, energy or infrastructure. Those projects can have significant social impacts in a territory and raise various political or economic issues. Understanding the human geography and recognizing the interests of the communities cohabiting in that territory can be critical to obtain a buy-in from all stakeholders. Onomastics, combined with geo-demographic segmentation, can help rapidly build geographic maps that can be used both for decision making and communication purposes. Automatic name clustering is the underlying technology that will help decrypt the complex identities present in large or small territories (from a continent, to a road). The objective is to answer tough questions and manage unavoidable frustrations though appropriate communication. Where should a tramway line pass in a multi-ethnic region? How to redistribute offshore oil revenues in the lands?

Concerning HR, I recently spoke with an executive at a large European bank who regretted that not enough trustworthy expatriates had been sent to control a large acquisition in a BRIC country, costing several hundred million Euros in write-offs. Among thousands of employees at the European head-office, the bank could have recognized the names of few people likely to accept an expatriation back to their home country. Having some people knowing both languages and both corporate cultures would have helped bridge the inter-cultural gap between the local management and other expatriates, saving millions of Euros.

In the digital world, onomastics brings a new view angle to social graph analysis: it can help colourize online communities, profile opinion leaders according to their audience. On Twitter, for example, you can more easily create a communication channels, well targeted on a particular community (business expatriates, tourists, migrants, but also international investors…)

Let's now consider a provocative and controversial use of onomastics that will help us move on to the topic of ethics. Different cultures, nationalities and social backgrounds imply different behaviours, with respect to Money and Risk taking: earning, saving, spending, gambling, investing, donating, risking death and loosing it all… It is a fact that people with aristocratic names (in places where there is such an object as aristocratic names) would earn more and obtain cheaper credit than people with names typical of the lower class or a recent immigration wave. Why not take shortcuts: a bank could adjust the price of a credit, according to the borrower's name; a car insurance company could adjust its evaluation of the risk (including the risks of insurance fraud, dangerous driving…) according to the name on the application form. They would better measure their risk. Furthermore, they could offer more competitive prices for categories of clients and they could better target them commercially.

Such use is highly controversial, since it raises the question of Equality (or inequalities) and discrimination. But discrimination is a fact, and onomastics can allow us to better see and understand how it works. Why should people with different sounding names hit glass ceilings in the first place, regardless of their skills? Casanova chose his own name de Seingalt and wondered if D'Alembert would have attained his high fame, his universal reputation, if he had been satisfied with his name of M. Le Rond, or Mr. Allround.

I am a supporter of Equal Opportunity Rights. And yet, I built a powerful discrimination algorithm based on names. NamSor is a piece of name recognition software which applies onomastics to analyse global flows of money, ideas and people. As any powerful new technology, it carries potential risks of abuse but I believe there is a positive use for it.

One classical application where onomastics plays a significant role is called geo-demographics: it consists in analysing the sociology of a particular territory (including the cultural and ethnic origins of its inhabitants) inferred from open sources and census data. Geo-demographics can be a useful tool to ensure, for example, that all populations have an equitable access to public services, such as hospitals. The company Experian is one of the leaders in that field, especially strong in the UK.

The effective use of the Big Data & Open Data is widely considered to be a critical enabler for future SmartCities : enabling dynamic allocation of resources, more efficient use of energy, prompt response to a crisis and so on. The combination of social networks and mobile applications with geo-localized devices opens new possibilities. Recognizing the diversity of populations that cohabit across space and time can help design more inclusive cities and transportation systems. Sensors that discriminate populations (in the sense of perceiving) can draw the clear picture needed to prevent discrimination (in the sense of favouring) and help defuse some of the time bombs ticking here and there.

**Targeting diasporas: a game changer for development?**
But the most promising use of software such as NamSor could be elsewhere – though it still deals with territorial equality. It is quite common for regions of the world that are less economically developed to use their own weakness (poorer people) as a strength (cheaper labour) to attract investments. The idea is to trigger a virtuous circle of job creation, infrastructure development, better education, migration flow reversal, etc. commonly known as the FDI Magnet effect. The region becomes more attractive and gradually moves up in the global value chain. As it loses competitiveness in terms of cheap labour because of the new wealth of its population, it develops a different economy based on innovation, services, tourism, consumption.

Most countries implement some kind of policy to direct flows of investments in poorer regions, as a mean to preserve their territorial cohesion and integrity. Those policies are most effective when they combine with successful private initiatives. So the objective of many Investment Promotion Agencies (IPA) is not so much to attract big money, as to attract a great business that will employ and help grow their people. The global competition to attract such investments is fearful.

Poorer regions have another weakness, which can be turned into a strength. Emigration is generally an opportunity loss, but after some years it generates a Diaspora which can be leveraged to attract investments back to the region.

For example, Ireland took decisive steps during the early 80's to proactively reconnect with its emigrants or with successful businessmen of Irish descent. Rebekah Berry reminds us that "as recently as 1986 Ireland was one of the poorest countries in the European Union, but [in 2002] it is one of the richest. The engine of this new Irish prosperity has been Foreign Direct Investment (FDI). [Between 1986 and 2002], the Irish have done almost everything right. They have attracted huge amounts of money from America – due largely to a century of personal and familial ties – and they have used this money to build factories ".

The regions of Ningxia, Gansu and Qinghai have amongst the lowest number of millionaires in China. But if they could reconnect with the few they have, in Beijing, Shanghai or even abroad, wouldn't it make a difference?

For that purpose, onomastics can be a useful tool and it has served the development strategy of a European country, Lithuania.

InvestLithuania is the first Investment Promotion Agency (IPA) to use name recognition to originate FDI deals. With three million people living in Lithuania and nearly one million people of Lithuanian origin living abroad, there is a good many personal and familial ties to be leveraged to attract new investment projects to the country. NamSor name recognition software helped discover those ties. Another method to accelerate the origination of new investment leads is to better understand and leverage the existing network of foreign businessmen in the country itself. Domas Girtavicius, a Senior consultant at Invest Lithuania, said "we were impressed by the accuracy of the name recognition software: it reliably predicts the country of origin and the number of false positives is fully manageable".

This project with InvestLithuania was very successful and consequently I was invited to participate to the World Lithuanian Economic Forum (WLEF), which took place in Vilnius this year, on the 3rd of June. This Forum is organized by Global Lithuanian Leaders (GLL), a non-profit association whose mission is to reconnect with Lithuanians and friends of Lithuania abroad. I found the GLL to be a great initiative, providing the country with a wealth of expertise from different parts of the word, across all domains (politics, education, culture, business…), and also bridging some of the cultural gaps that necessarily exist in such a matrix (place / domain). Specifically, the GLL helps bring elements of culture from the US and UK, such as entrepreneurship and business networking.

While some diasporas, especially those originating from the Mediterranean, have a millennium standing culture of business and personal networking, other countries struggle to adjust to their new situation. What is the value of a social network such as LinkedIn to the Lebanese Diaspora? Low. What better communication tool in Marseilles than "word of mouth" to launch Massilia Mundi, which aims to become the social network of that city international Diaspora? But for many Investment Promotion Agencies (IPAs), LinkedIn is an essential tool. For example, in traditional Lithuanian culture, people treasure strong family ties and personal links with close friends, but do not nurture a wide network of professional connections or casual contacts. I believe many countries are in a similar situation, where a dedicated organisation could help reconnect people : for them, tools such as the social networks, professional databases and onomastics can make a difference.

Could that work also for regions in China? In 2005-2009, while I was working for a global consulting firm, I had the opportunity of managing an project in banking, with a mix of Chinese and French teams: a team in Paris which included several young ParisTech graduates of Chinese origin and a team in Shanghai. I remember the excitement and the pleasure of the entire team – including myself – to do a project connected with China, with the opportunity of travelling to Shanghai, tasting the food of different regions of China, being introduced to the Chinese culture. Several people from that team, both French and Chinese, are now in China. Jing, now a dear friend, went back to Shanghai in 2009 and I remember how she still felt sentimentally bound to her original city of Xiangtan, Hunan – ready to help in any way she could. From this experience, I understood that if there existed such an organization as 'Global Ningxia, Gansu and Qinghai Leaders, it would not often encounter rebukes when reaching out for help, money or expertise. Such an organization could be very helpful in closing the economic gap with other regions.

Technically, Chinese names are clearly recognizable amongst other nationalities or origins. So, querying a professional database, we can produce onomastics mapping of Chinese company directors. For example, the following maps represents the density of Chinese and Japanese business communities in Southern Latin America, relatively to each other.

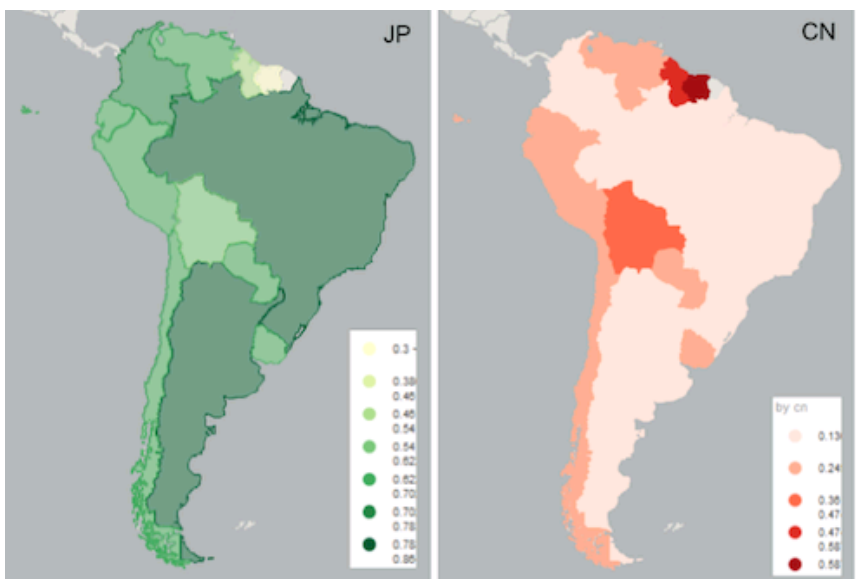

Source: Factiva DF Copyright 2013 NamSorts.com NomTriTM NamSorTM – All rights

How many of those successful Chinese businessmen (or businessmen of Chinese origin) come from Ningxia, Gansu or Qinghai? This is where applied onomastics can be a game changer. Not that all questions are solved. At the present time, the available software allows us to detect phenomenons, not to understand them perfectly. For instance, I would like to share two data visualizations produced as part of this effort, which I found beautiful and promising.

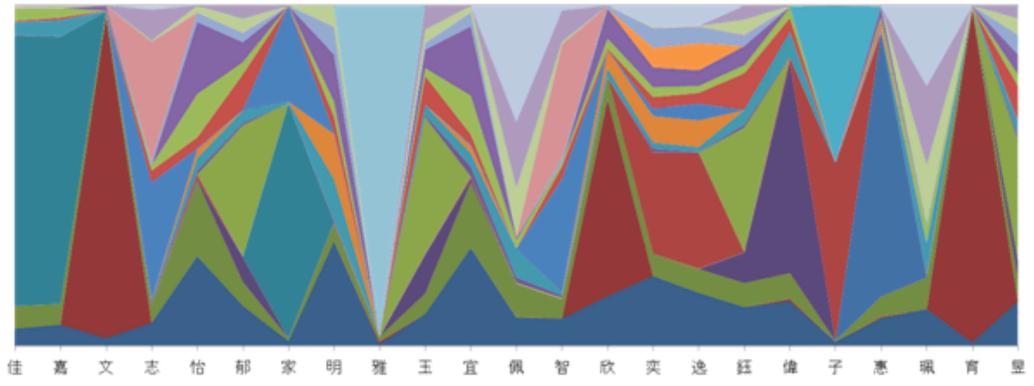

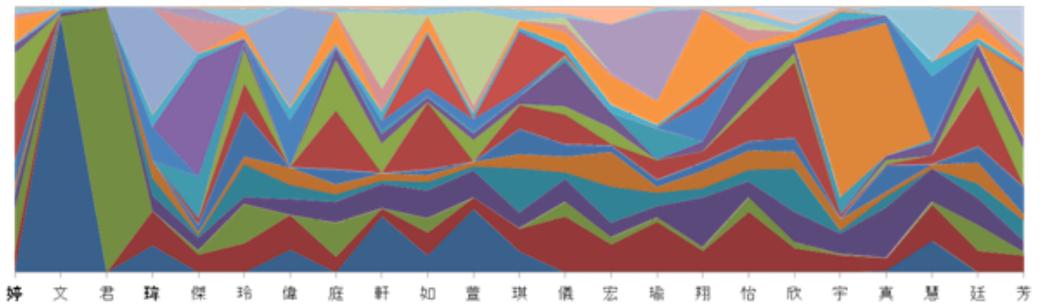

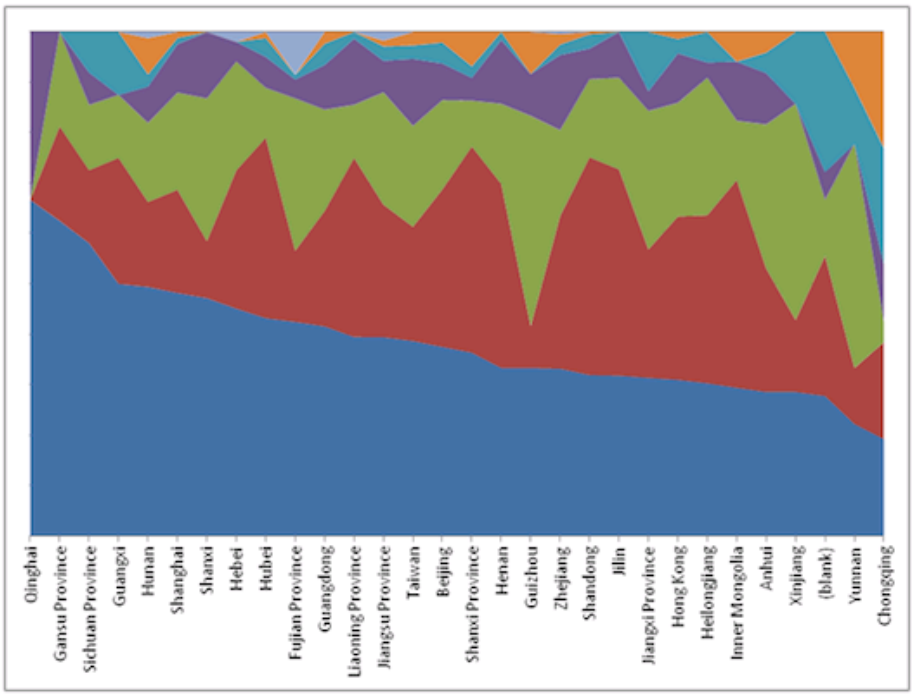

What do we see here? Something – something that still needs to be analysed and understood, but something that may be of great value for someone trying to locate and identify potential investors or decision makers. Chinese last names actually raise specific challenges, since they have been used for many centuries and with rare or less common names disappearing over time, only one hundred different names remain today. But first names still carry regional differences, poetry and other semantics. Roots may be almost invisible, onomastics can still track them. And the more difficult is the tracking, the more valuable are the findings.

More on paristech review
On the topic
IT for ID: India's Biometrics MegaprojectBy Nandan Nilekani on January 31st, 2013
Think Big: Challenges and Debates over the Big DataBy Henri Verdier on November 16th, 2012
The Allure of Big DataBy Jannis Kallinikos on November 16th, 2012
Have New Advertising Models Become Too Aggressive and Intrusive?By Knowledge@Wharton on April 27th, 2012
Turning the Data Deluge into Competitive AdvantageBy Knowledge@Wharton on November 25th, 2011
Does the Rise of Surveillance Mean the End of Privacy?By ParisTech Review on September 15th, 2010
Voice on the Border: Do Cellphones Redraw the Maps?By Vincent Blondel, Pierre Deville, Frédéric Morlot, Zbigniew Smoreda, Paul Van Dooren & Cezary Ziemlicki on November 15th, 2011
By the author
Onomastics for Business: can discrimination help development?on September 17th, 2013

ParisTech REVIEW: